\documentclass[10pt,b5paper,twoside]{padeu}  
\usepackage{epsfig,natbib_padeu}              
\usepackage{amsmath}

\begin{document}

\title{Generalized eccentric vs. true anomaly parametrizations in 
the perturbed Keplerian motion}
\titlerunning{Generalized eccentric vs. true anomaly}
\author{Bal\'{a}zs Mik\'{o}czi\inst{1},
        Zolt\'{a}n Keresztes\inst{2}}
\authorrunning{B. Mik\'{o}czi, Z. Keresztes}
\institute{Departments of Theoretical and Experimental Physics, 
University of Szeged, 6720 Szeged, D\'{o}m t\'{e}r 9, Hungary}
\email{\inst{1}mikoczi@titan.physx.u-szeged.hu,        
       \inst{2}zkeresztes@titan.physx.u-szeged.hu}


\abstract{
The angular and the radial parts of the dynamics of the perturbed 
Kepler motion 
are separable in many important cases. In this paper we study the 
radial motion and its parametrizations. 
We develop in detail a generalized eccentric anomaly 
parametrization and a 
procedure of computing a generic class of integrals based on the 
residue theorem.
We apply the formalism to determine various contributions to the 
luminosity of a compact binary.
}

\maketitle

\section{Introduction}

The main sources of the gravitational radiation to be detected by 
Earth-based observatories (LIGO, VIRGO, TAMA and GEO) are compact 
binaries. The Laser Interferometer Space Antenna (LISA) will also 
detect gravitational waves from colliding galactic black holes. 
Compact binaries are composed of black holes and/or neutron 
stars. In such systems, a highly accurate description would include 
the spin-spin (SS) interaction [\citet{Barker}], the magnetic
dipole-magnetic dipole moments (DD) contribution [\citet{Ioka}] 
and the quadrupole-monopole effect (QM) [\citet{Poisson}]. 
The parametrization of the radial motion for them are presented 
in [\citet{Gergely1}], [\citet{VKMG2003}], and [\citet{GK2003}].

The method for solving a wide class of the perturbed Keplerian radial
motions is worked out in [\citet{GPV2000}]. For this purpose two 
(the true and eccentric anomaly) parametrization were introduced. 
The method can be further generalized for an even wider class of 
perturbed Keplerian radial motions (see: [\citet{GKM2006}]).

\section{The perturbed Kepler motion}

The so-called radial equations for SS, DD and QM interactions are 
given in [\citet{KMG2005}]. The radial equation of the
generalized perturbed Kepler motion we consider here [\citet{GKM2006}] is
\begin{equation}
\dot{r}^{2}=\frac{2E}{\mu }+\frac{2Gm}{r}-\frac{L^{2}}{\mu 
^{2}r^{2}}
+\sum_{i=0}^{p}\frac{\varphi _{i}(\chi )}{\mu ^{2}r^{i}}\ \,   
\label{radial}
\end{equation}
where $\varphi _{i}(\chi )$ characterize the perturbative terms and 
they are periodic functions of the true anomaly: 
\begin{equation}
\varphi _{i}\left( \chi \right) =\sum_{j=0}^{\infty }\left(
f_{ij}+g_{ij}\cos \chi \right) \sin ^{j}\chi \ .  \label{phi}
\end{equation}
The expression $\varphi _{i}(\chi )$ given above is equivalent 
with a generic Fourier series [\citet{GKM2006}]. The coefficients 
$f_{ij}$ and $g_{ij}$ can be expressed in terms of the coefficients of the
Fourier expansion as well. The last term in Eq. (\ref{radial}) contains 
the generic perturbing Brumberg force [\citet{Brumberg}], 
the spin-orbit interaction for compact binaries [\citet{RI1997}; \citet{GPV1998}], 
and the SS, DD and QM contributions. The energy $E$ and angular momentum $
L $ refer to the perturbed motion. From the condition $\dot{r}^{2}=0$ we find 
the turning points $r_{{}_{{}_{\min }^{\max }}}$:
\begin{equation}
r_{{}_{{}_{\min }^{\max }}}=\frac{Gm\mu \pm A_{0}}{-2E}\pm 
\frac{1}{2\mu
A_{0}}\sum_{i=0}^{p}\varphi _{i}^{\pm }(\chi )\left[ \frac{\mu 
(Gm\mu \mp
A_{0})}{L^{2}}\right] ^{i-2}\ ,  \label{rminmax}
\end{equation}
where $A_{0}=\left( G^{2}m^{2}\mu ^{2}+2EL^{2}/\mu \right) 
^{1/2}$ is the
magnitude of the Laplace-Runge-Lenz vector belonging to the 
perturbed motion characterized by $E$ and $L$. $\varphi 
_{i}^{-}=\varphi
_{i}(0)$, $\varphi _{i}^{+}=\varphi _{i}(\pi )\ $are small 
coefficients.
With these turning points is possible to introduce the generalized
true anomaly parametrization of the radial motion, $r\left( \chi \right) 
$ as
\begin{equation}
\frac{2}{r}=\frac{1+\cos \chi }{r_{\min }}+\frac{1-\cos \chi 
}{r_{\max }}\ .
\label{true1}
\end{equation}

Integrals of the type 
\begin{equation}
\int_{0}^{T}\frac{\omega (\chi )}{r^{2+n}}dt\  \label{int}
\end{equation}
frequently occur. Here $\omega (\chi )$ is the same type of periodic 
function of the true anomaly as $\varphi _{i}\left( \chi \right) $, 
and $T$ \ is the radial period of the motion. 

For constant values of $\varphi _{i}$ and $n\geq 0$ 
these integrals can be evaluated in terms of a  
complex true anomaly variable $z=\exp \left( i\chi 
\right) $ by using the residue theorem. The only pole is in the origin ($z=0$) 
[\citet{GPV2000}]. For the class $n<0$ a complex eccentric anomaly parameter 
can be used and then the poles are in the origin and in 
\begin{equation}
w_{1}=\left( \frac{Gm\mu ^{2}-\sqrt{-2\mu EL^{2}}}{Gm\mu ^{2}+\sqrt{-2\mu
EL^{2}}}\right) ^{1/2}\ ,
\end{equation} 
however the latter occurs only rarely in physical applications 
(see:[\citet{GPV2000}]).

\section{Generalized perturbed Kepler motion with the eccentric 
anomaly parametrization}

We introduce the $r\left( \xi \right) $ eccentric anomaly
parametrization in the same way as in [\citet{GPV2000}]:
\begin{equation}
{2}r=(1+\cos \xi )r{_{\min }}+(1-\cos \xi )r{_{\max }}\ .  
\label{eccentric1}
\end{equation}
We use the relations between the true and eccentric anomaly 
(\ref{true1}), (\ref{eccentric1}) 
\begin{equation}
\cos \chi =\frac{Gm\mu \cos \xi -A_{0}}{Gm\mu -A_{0}\cos \xi }\ 
,\qquad \sin
\chi =\frac{\sqrt{\frac{-2EL^{2}}{\mu }}\sin \xi }{Gm\mu 
-A_{0}\cos \xi }\ .
\end{equation}
Thus we can express $\varphi _{i}$ as the function of $\xi$:
\begin{equation}
\varphi _{i}\left( \xi \right) =\sum_{j=0}^{\infty }\left( 
f_{ij}+g_{ij}
\frac{Gm\mu \cos \xi -A_{0}}{Gm\mu -A_{0}\cos \xi }\right) \left( 
\frac{
-2EL^{2}\sin \xi }{Gm\mu ^{2}-A_{0}\mu \cos \xi }\right) ^{j}\ .
\label{phi xi}
\end{equation}
Employing the eccentric anomaly parametrization 
(\ref{eccentric1}), the
integrals (\ref{int}) could be evaluated as
\begin{equation}
\int_{0}^{2\pi }\frac{\omega (\xi )}{r^{n+1}}\left( 
\frac{1}{r}\frac{dt}{
d\xi }\right) d\xi \ .  \label{int2}
\end{equation}
For $n^{\prime }\equiv -n-1\geq 0$ we apply the 
binomial expansion 
\begin{equation}
\left( {2}r\right) ^{n^{\prime }}=\sum_{k=0}^{n^{\prime }}\left( 
_{n^{\prime
}}^{k}\right) r_{\min }^{k}r_{\max }^{n^{\prime }-k}(1+\cos \xi 
)^{k}(1-\cos
\xi )^{n^{\prime }-k}\ ,
\end{equation}
leading to a polynomial in $\cos \xi $. From the radial equation 
(\ref{radial})
using the eccentric anomaly parametrization (\ref{eccentric1}) to
leading order we obtain:
\begin{equation}
\frac{1}{r}\frac{dt}{d\xi }=\sqrt{\frac{\mu }{-2E}}\biggl 
[1-\frac{E}{2\mu
A_{0}^{2}\sin ^{2}\xi }\sum_{i=0}^{p}\left( \Omega 
_{+}^{i}-\Omega
_{-}^{i}\cos \xi -\frac{2\varphi _{i}\left( \xi \right) 
}{r^{i-2}}\right) 
\biggr ]\ ,  \label{dtdxi}
\end{equation}
with
\begin{equation}
\Omega _{\pm }^{i}\!=\!\left( \frac{\mu }{L^{2}}\right) 
^{i-2}\!\left[
\varphi _{i}^{+}(Gm\mu \!-\!A_{0})^{i-2}\!\pm \!\varphi 
_{i}^{-}(Gm\mu
\!+\!A_{0})^{i-2}\!\right] \ .
\end{equation}
The bracket becomes proportional to $\sin ^{2}\xi $ in 
(\ref{dtdxi}) if the following two 
conditions are satisfied: 
\begin{equation}
\sum_{i=0}^{p}\frac{\mu ^{i}}{L^{2i}}\left( Gm\mu \pm 
A_{0}\right)
^{i}\left( f_{i1}\pm g_{i1}\right) =0\ .  \label{cond}
\end{equation}

With conditions (\ref{cond}) satisfied, using the parametrization 
(\ref{eccentric1}), the
integrand of (\ref{int2}) becomes regular. The conditions are 
fulfilled in the case of the
SS, DD and QM interactions. Introducing the complex variable 
$w=\exp \left( i\xi \right) $, the integral of (\ref{int2}) 
contains only two poles: the origin and the
$w_{1}$ (see:[\citet{GPV2000}]). We have proven for the 
$n<0$ case:

\textbf{Theorem}: \textsl{For all perturbed Kepler motions 
characterized by
the radial equation (\ref{radial}), with periodic perturbing 
functions }$
\varphi _{i}(\xi )$\textsl{\ obeying the conditions 
(\ref{cond}),\ and for
arbitrary periodic functions }$\omega (\xi )$,\textsl{\ the 
integrals (\ref
{int2}) } \textsl{are given by the residue in the origin and in 
the }$
w_{1}$\textsl{\ [\citet{GPV2000}]. on $w$ complex plane.}

In the next section we apply the above method for computing 
different contributions to the luminosity of compact binaries.

\section{Application for some compact binaries}

Peters and Mathews [\citet{PM1963}] computed the luminosity of 
the compact binaries in
the Kepler motion (Newtonian case) with orbital parameters:
\begin{equation}
\mathcal{L}_{N}=\frac{G^{4}m^{3}\mu 
^{2}}{15c^{5}a^{5}(1-e^{2})^{7/2}}\left(
37e^{4}+292e^{2}+96\right)\ ,
\end{equation}
where $a$ the semi-major axis and $e$ the eccentricity. The form of 
$\varphi _{i}\left( \xi \right) $ can be derived for all of the SS, QM and 
DD contributions. We can then compute the various contributions
 to the luminosity ($-\left\langle dE/dt\right\rangle$, the time-averaged 
energy loss over one radial period). These are:
\begin{eqnarray}
\mathcal{L}_{S_{1}S_{1}}\! &=&\frac{G^{4}m^{2}\mu S_{1}S_{2}}{
480c^{7}a^{7}(1-e^{2})^{10/2}}\!\biggl [c_{1}\sin \kappa _{1}\sin 
\kappa
_{2}\cos 2(\psi _{0}-\bar{\psi})  \notag \\
&&+c_{2}\cos \kappa _{1}\cos \kappa _{2}+c_{3}\cos \gamma \biggr 
]\ ,  \notag
\\
\mathcal{L}_{QM} &=&\!\frac{G^{4}m^{5}\mu 
}{60c^{5}a^{7}(1-e^{2})^{10/2}} 
\notag \\
&&\times \sum_{i=1}^{2}p_{i}\left[ \left( c_{4}(3\sin ^{2}\kappa
_{i}-2)+c_{5}\sin ^{2}\kappa _{i}\cos \delta _{i}\right) \right] 
\ ,  \notag
\\
\mathcal{L}_{DD} &=&\frac{G^{3}m^{2}\mu d_{1}d_{2}}{
30c^{5}a^{7}(1-e^{2})^{10/2}}\!\left[ 
c_{4}\mathcal{A}_{0}-c_{5}\mathcal{B}
_{0}\right] \ ,
\end{eqnarray}
where we have denoted by $\ S_{i}$ the spin magnitudes, by $d_{i}$ the magnitudes of the magetic dipoles, by $p_{i}$ the quadrupole moment scalars and by $\mathcal{A}_{0}$, $\mathcal{B}_{0}$, $\delta_{i}$, $\gamma $, $\psi _{0}$, $\bar{\psi}$, $\kappa _{i}$ 
auxiliary angular quantities defined in [\citet{Gergely1}], [\citet{VKMG2003}] and 
[\citet{GK2003}]. The constants $c_{1..5}$ are
\begin{equation}
c_{i}=\sum_{j=0}^{3}C_{ij}e^{2j}\ ,
\end{equation}
with coefficients $C_{ij}$ given in Tab.~\ref{Table1}.

\begin{table}
   \caption{The coefficients in the $C_{ij}$.}
   \label{Table1}
   \begin{center}
   \begin{tabular}{ccccc}
   \hline
   \hline
   \noalign{\smallskip}
    $_{i}\backslash ^{j}$ & 0 & 1 & 2 & 3 \\
   \noalign{\smallskip}
   \hline
   \noalign{\smallskip}
   1 & 0 & 131344 & 127888 & 7593 \\ 
      2 & -124864 & -450656 & -215544 & -8532 \\ 
      3 & 42048 & 154272 & 75528 & 3084 \\ 
      4 & 0 & 8208 & 7988 & 474 \\ 
      5 & 2600 & 9376 & 4479 & -177 \\
   \noalign{\smallskip}
   \hline
   \noalign{\smallskip}
   \end{tabular}
   \end{center}
   \end{table}

\section{Summary}
We have introduced a generalized eccentric anomaly parametrization 
for the perturbed Kepler motion. We have proved that even for the 
generic perturbation functions $\varphi_i(\xi)$ there are no new poles 
as compared to [\citet{GPV2000}]. The method of integration can be 
widely employed in the case of compact binaries.
We have applied the procedure to compute the SS, QM and DD contributions 
to the luminosity of a compact binary. 

\section{Acknowledgements}

This work was supported by OTKA grants no. T046939 and TS044665. 
We thank L. \'{A}. Gergely for guidance in the topic.

\end{document}